\documentclass[aip,reprint,onecolumn]{revtex4-1}

\usepackage[pdftex]{graphicx}% Include figure files
\usepackage{dcolumn}% Align table columns on decimal point
\usepackage{bm}
\usepackage{verbatim}
\usepackage{dsfont}
\usepackage{mathrsfs}
\usepackage{lipsum}
\usepackage{geometry}
\usepackage{color}
\usepackage{gensymb}

\begin{document}

%\preprint{}

\title{A versatile, inexpensive integrated photonics platform} 

\author{Jeffrey M. Shainline}
\email[]{jeffrey.shainline@nist.gov}

\author{Sonia M. Buckley}
\affiliation{National Institute of Standards and Technology, 325 Broadway, Boulder, CO, 80305, USA}

\author{Nima Nader}
\affiliation{National Institute of Standards and Technology, 325 Broadway, Boulder, CO, 80305, USA}

\author{Cale M. Gentry}
\affiliation{Department of Electrical, Computer and Energy Engineering, University of Colorado, Boulder, CO, 80309, USA}

\author{Kevin C. Cossel}
\affiliation{National Institute of Standards and Technology, 325 Broadway, Boulder, CO, 80305, USA}

\author{Milo\v{s} Popovi\'{c}}
\affiliation{Department of Electrical, Computer and Energy Engineering, University of Colorado, Boulder, CO, 80309, USA}

\author{Nathan R. Newbury}
\author{Sae Woo Nam}
\author{Richard P. Mirin}
\affiliation{National Institute of Standards and Technology, 325 Broadway, Boulder, CO, 80305, USA}

%\thanks{}

\date{\today}

\begin{abstract}
We present an approach to fabrication and packaging of integrated photonic devices that utilizes waveguide and detector layers deposited at near-ambient temperature. All lithography is performed with a 365 nm i-line stepper, facilitating low cost and high scalability. We have shown low-loss SiN waveguides, high-$Q$ ring resonators, critically coupled ring resonators, 50/50 beam splitters, Mach-Zehnder interferometers (MZIs) and a process-agnostic fiber packaging scheme. We have further explored the utility of this process for applications in nonlinear optics and quantum photonics. We demonstrate spectral tailoring and octave-spanning supercontinuum generation as well as the integration of superconducting nanowire single photon detectors with MZIs and channel-dropping filters. The packaging approach is suitable for operation up to 160 \degree C as well as below 1 K. The process is well suited for augmentation of existing foundry capabilities or as a stand-alone process.
\end{abstract}

%\pacs{85.25.Am,85.25.Cp,85.25.Dq,85.25.Hv,85.25.Oj,85.25.Pb,85.60.Dw,85.60.Jb}

\maketitle

\section{\label{sec:intro}Introduction}		
The first on-chip photonic components were developed in the mid-1980s, and the subsequent three decades have demonstrated exponential growth in the data capacity per die \cite{kiwe2011} due to the miniaturization enabled by integrated photonic devices. Photonic integrated circuits are now playing a role in advanced technologies with applications ranging from atmospheric spectroscopy \cite{naco2016} and medical diagnostics \cite{voar2008,joro2009,wagu2009,walu2010,chyu2015} to telecommunications \cite{yats2014,do2015}, data centers, \cite{kato2012,niru2015} and supercomputers \cite{runi2015,suwa2015}. Each of these integrated photonic application spaces opens new possibilites for science and technology. 

The chip-scale photonic devices enabling this innovation are manufactured using lithographic techniques developed by the semiconductor industry, and it is the precision, accuracy, and reliability of these fabrication techniques that has enabled the rapid adoption of integrated photonics for myriad applications. However, the broad range of integrated photonic technologies and application spaces makes it more difficult to create a technology roadmap and a foundry process with broad applicability than it has been for electronics. %Still, significant progress is being made both in device innovation and large-scale manufacturing.  

The American Institute for Manufacturing of Integrated Photonics (AIM Photonics) \cite{aim} was launched in 2015, moving integrated photonics closer to the foundry model that has been so successful for integrated electronics. AIM Photonics intends to provide  multi-project wafer (MPW) runs, enabling start-ups, small businesses, and academics to leverage the established manufacturing infrastructure. However, many application spaces will not be served by the specific fabrication processes developed for the AIM Photonics technology focus areas. This is unavoidable in the near future due to the enormous diversity of applications of integrated photonic devices and systems. For nonlinear optics in the visible and near-infrared, Si is not a desirable waveguiding medium due to high one- and two-photon absorption \cite{Moss2013}. For nonlinear optics in the mid-IR, the absorption of a standard buried SiO$_2$ layer is problematic \cite{Soref2006}. For quantum-optical applications such as quantum networks, quantum computers, and quantum metrology, high detection efficiency is required, and superconducting detectors are a very desirable option \cite{Marsili2013}. Such detectors are not likely to be incorporated into a foundry process in the near term.  

In the present work, we show how a variety of alternative application spaces can be served with a simple, low-cost approach to fabrication of integrated photonic devices and systems as well as a simple back-end-of-line (BEOL) approach to fiber packaging. We introduce a modular fabrication process, employing layers of dielectrics, superconductors, metals, and polymers deposited at near-ambient temperatures ($<$ 65 \degree C). The dielectrics are deposited using plasma-enhanced chemical vapor deposition (PECVD), and the superconductors are deposited with sputtering. Such deposition techniques are compatible with BEOL processing, so these process modules and photonic devices can be straightforwardly integrated with fully processed wafers from other foundries. For BEOL integration with CMOS electronics, deposition temperatures of less than 400 $^\circ$C can meet many process requirements, and several photonic devices have been demonstrated in such a process \cite{Lee2013}. Room temperature deposited materials \cite{shch2016} are even more broadly compatible. With this fabrication process, we have demonstrated low-loss passive SiN waveguides and resonators, critically coupled rings, and Mach-Zehnder interferometers (MZIs), dispersion-engineered waveguides and waveguide-device-integrated WSi superconducting-nanowire single-photon detectors (SNSPDs). 

\section{\label{sec:basicPassives}Fabrication and passive devices}
\begin{figure} %[t] %[htb]
	\includegraphics{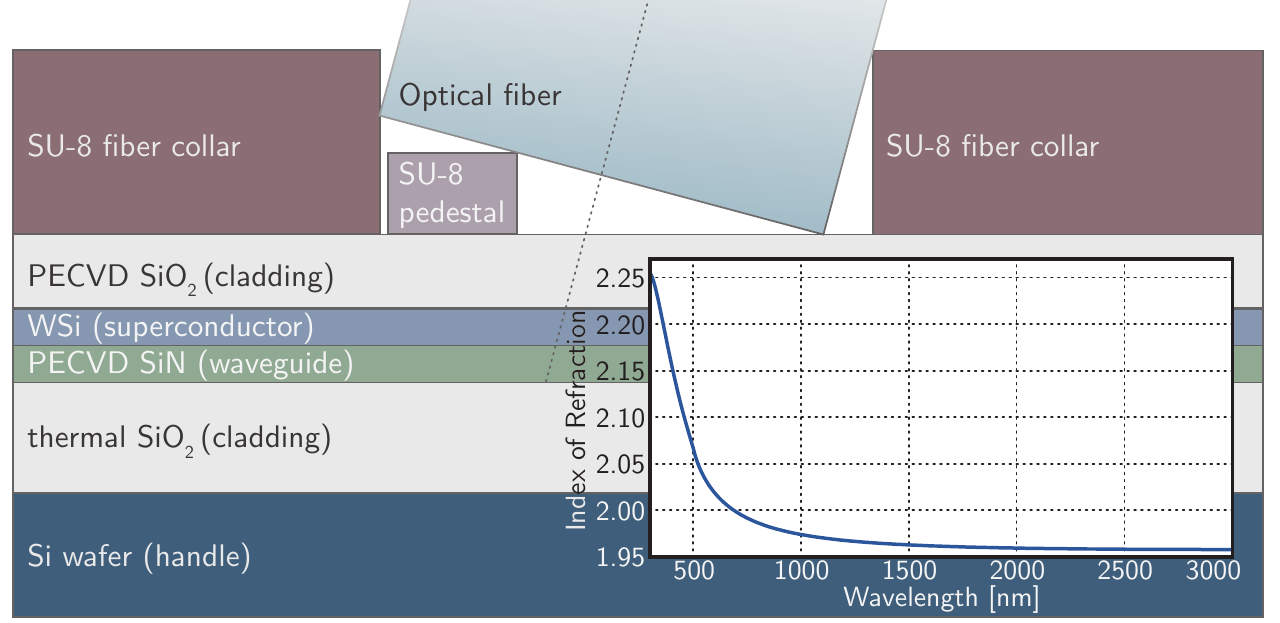}
	\caption{\label{fig:layers} A schematic of the layers used in this fabrication process. The inset shows the index of refraction of the SiN waveguide layer as measured with spectroscopic ellipsometry.}
\end{figure}

While most Si photonics foundry processes to date utilize crystalline Si as a waveguiding layer, deposited amorphous materials offer several advantages such as low-propagation-loss amorphous silicon (a-Si) \cite{zhlo2010,tama2014} as well as SiN waveguides \cite{sayo2016,shch2016}, a wide variety of indices of refraction, lower loss and crosstalk in waveguide crossings \cite{sayo2016}, and 3D layering \cite{sayo2016,Biberman2011,Sherwood-Droz2011,Bauters2011}. A schematic of the layers used in this process is shown in Fig. \ref{fig:layers}. The process begins with a handle Si wafer on which 3 $\mu$m of thermal SiO$_2$ is grown as a bottom cladding layer. Thermal SiO$_2$ could easily be replaced by PECVD SiO$_2$ deposited at 65 \degree C\ for compatibility with BEOL processing. Alternatively, the fabrication process could commence with a handle wafer other than oxidized Si such as sapphire, quartz, diamond, or III-V materials. Next, the SiN waveguiding layer is deposited using plasma-enhanced chemical vapor deposition\cite{shch2016} using N$_2$ and SiH$_4$ as precursors (27 and 30 sccm respectively). The plasma is struck at 30 W and cut to 0 W after 10 s. The chamber pressure is 10 mTorr, and the inductively-coupled power is 1000 W. The wafer temperature has been monitored and does not exceed 65 \degree C. The resulting SiN film has low stress due to high hydrogen content, so films as thick as 900 nm have been deposited with no cracking or delamination. Even with the excess hydrogen, low propagation losses are observed, as will be discussed shortly. 

Following the SiN deposition, and without breaking the vacuum of the PECVD chamber, a thin spacer SiO$_2$ layer is deposited as an etch stop for the WSi SNSPD etch. The wafer is removed from the PECVD chamber and transferred to a sputtering chamber where 3.5 nm WSi is deposited followed by a 2 nm amorphous Si layer \cite{Marsili2013}. Both of these sputtering steps are performed at ambient temperature. Alignment marks are etched in the stack. The SNSPD layer is next patterned using using 365 nm i-line stepper photolithography to achieve wires as narrow as 250 nm. The SNSPDs are etched using 40 sccm Ar and 1 sccm SF$_6$ to minimize sidewall residue. The photoresist is stripped, and the waveguide layer is patterned, again using the i-line stepper. The SiN waveguides and all photonic devices (rings, beam splitters, grating couplers) are etched using CF$_4$ chemistry. 

After patterning of the SNSPDs and waveguides, Au pads with a Ti adhesion layer are deposited to make contacts to the SNSPDs. Subsequently, the entire wafer is clad with 2 $\mu$m of PECVD SiO$_2$ deposited with the same tool as was used for the SiN waveguiding layer. Following the cladding deposition, vias are etched through this cladding to allow access to the Au pads for wire bonding. The final layers are SU-8 polymer layers used for packaging and will be discussed in more detail in Sec. \ref{sec:packaging}. Fig. \ref{fig:layers} shows the full layer stackup.

\begin{figure} %[t] %[htb]
	\includegraphics{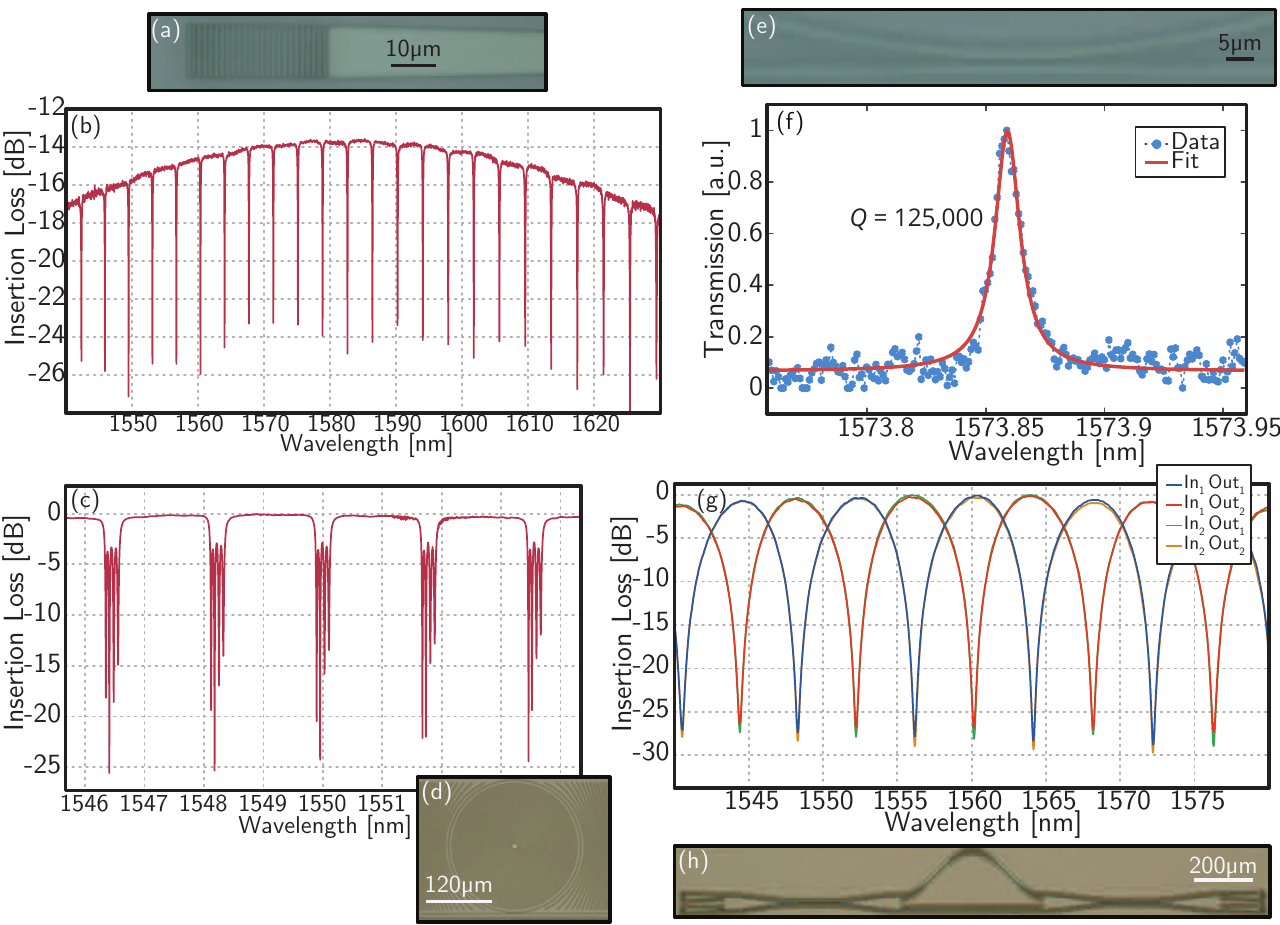}
	\caption{\label{fig:passives} (a) Microscope image of a fabricated grating. The period is 1.4 $\mu$m, and the duty cycle is 60 \%. (b) Spectral response of a single ring with outer radius of 60 $\mu$m, width of 1.5 $\mu$m coupled to a waveguide between two gratings. The coupling bus has a width of 1.5 $\mu$m, and the ring-bus gap is 610 nm. (c) Spectrum of four rings achieving close to critical coupling with outer radii of 120 $\mu$m. (d) A single ring with small ring-bus gap for critical coupling. (e) Microscope image of a ring-bus coupler. (f) Fitted resonance for weakly coupled ring resonator. (g) Spectrum of the unbalanced MZI. (h) Optical microscope image of the MZI. (c) and (g) are normalized to the peak of the grating response.}
\end{figure}
To illustrate the utility of this process, we demonstrate several photonic devices. Figures \ref{fig:passives}(a) and (b) show microscope images of a SiN grating and the spectral response of a single ring coupled to a waveguide between two gratings. The insertion loss due to the gratings is 6.8 dB/coupler, in close agreement with the finite-difference time-domain simulated value of 6.68 dB/coupler. The 3 dB bandwidth of these couplers is 90 nm. This spectrum was taken after the fibers were fixed in place as described in Sec. \ref{sec:packaging}. While the lower index contrast of SiN limits grating efficiency, the insertion loss can be significantly improved with a thicker SiN layer \cite{Doerr2010}, a backside mirror \cite{Zhang2014}, a second deposited layer \cite{Roelkens2006}, or an additional (partial) etch step \cite{Maire2008}. With a backside mirror (which could be integrated easily in this process), insertion loss of 2.6 dB has been demonstrated \cite{Zhang2014}, and insertion loss of $<$ 2 dB has been simulated \cite{Doerr2010} in a SiN single-etch process similar to this work. The addition of more and higher index layers, such as a-Si, could also improve grating coupler insertion loss. State-of-the-art Si grating couplers have achieved insertion losses of less than 1 dB \cite{Mekis2011,Zaoui2014}, while grating couplers in a two-layer CMOS process with $\sim$ 0.2 dB coupling loss have recently been designed \cite{Notaros2015}. All of these approaches are compatible with the deposited photonics process and packaging approach presented here.

Next we investigated critically coupled ring resonators. In Fig. \ref{fig:passives}(c) we show the transmission spectrum of four rings coupled to a single bus. One such ring is shown in Fig. \ref{fig:passives}(d). Each of the four rings has a different gap ranging from 360 nm to 500 nm on the mask, corresponding to 610 nm to 750 nm after fabrication. The optimally coupled device achieved $>$ 25 dB extinction across several FSRs. To assess parasitic losses in the critically coupled rings, the most critically coupled resonance from each FSR of the spectrum in Fig. \ref{fig:passives}(c) was fit to a Lorentzian to obtain a $Q$ factor. The mean $Q$ of these resonances was 43,250 (corresponding to an intrinsic $Q$ of 86,500), and the standard deviation was 2,630.

In addition to critically coupled ring resonators, we have investigated weakly coupled, high-$Q$ rings. Figure \ref{fig:passives}(e) shows a microscope image of a ring-bus coupler, and Fig. \ref{fig:passives}(f) shows a fit to a weakly coupled resonance with $Q$-factor of 125,000, corresponding to 2.5 dB/cm propagation loss. The five resonances of this ring closest to 1570 nm were fit to obtain $Q$s. The mean $Q$ of the data set was 110,000 (corresponding to 2.8 dB/cm), and the standard deviation was 8,900. Sidewall roughness due to patterning with i-line lithography is likely the leading contribution to linear propagation losses, as lower losses have been observed in waveguides of similar material when patterned with electron beam lithography (0.8 dB/cm) [\onlinecite{shch2016}] as well as within this process when resist reflow was employed (1.8 dB/cm), as discussed in Sec. \ref{sec:nonlinear}. This is comparable to PECVD nitride deposited at 400$^\circ$C [\onlinecite{Lee2013}], but significantly higher than high temperature deposited LPCVD SiN, where losses of less than 0.1 dB/cm have been observed \cite{Bauters2011,Bauters2013,pfko2016,xuli2016}. Silicon waveguides typically exhibit propagation losses near 3 dB/cm [\onlinecite{Poon2015}]. Finally, in Fig. \ref{fig:passives}(g) we show the spectrum of an unbalanced MZI [Fig. \ref{fig:passives}(h)] demonstrating $>$ 25 dB extinction between all ports across $>$ 40 nm of spectral bandwidth. This extinction demonstrates that the adiabatic beam splitters created in our process are nearly ideal with no more than 1.5\% deviation from 50/50 splitting. If scattered light, noise from the laser source, or unbalanced losses are affecting the visibility, then the beam splitters are even closer to 50/50.  

%The ring-bus gap for the device from which this spectrum was acquired was 1.5 $\mu$m on the mask.

The thermal tuning of passive integrated-photonic devices is often advantageous. A process layer for heaters can be straightforwardly inserted either with evaporated metals or PECVD-deposited and doped a-Si. The a-Si option is particularly attractive due to the decreased electromigration relative to metals, and addition of a-Si to the process is also likely to lead to detectors for high-peak-power pulses used for nonlinear optics.

\section{\label{sec:packaging}Fiber Packaging}

Due to the absence of an integrated silicon light source, coupling external light on chip via a robust packaging technique has been an important technological goal for the integrated photonics community. Fiber packaging for operation at cryogenic temperatures is particularly challenging, and many researchers avoid packaging altogether by using fiber positioning stages in the cryostat\cite{Pernice2012b,Schuck2016,Sprengers2011,Najafi2015}. Alternatively, a high-numerical-aperture objective can be used to couple to waveguides through a window in the cryostat \cite{Akhlaghi2015}. Such systems typically achieve insertion losses $>$ 7 dB, although lower insertion losses have been reported \cite{Najafi2015}. Cryogenic coupling with tapered optical fibers has been achieved with losses below 1 dB \cite{srpa2007}. However, coupling with positioning stages requires frequent realignment and high-precision stages inside the cryostat where space is constrained. 

The fiber packaging approach presented here is achieved with two processing steps which construct fiber support structures above the grating couplers. These fiber collars comprise two layers of SU-8 epoxy-based photoresist which is permanent upon UV exposure and curing. First, a 20 $\mu$m pedestal layer is fabricated with SU-8 3020 photoresist. This layer serves to hold an optical fiber at the designed angle relative to the grating (15\degree\ in this work). The second SU-8 3050 layer is referred to as the collar, and it is 50 $\mu$m thick. The collar serves to align the fiber relative to the grating, and it provides a receptacle in which epoxy can be delivered.
\begin{figure} %[t] %[htb]
	\includegraphics{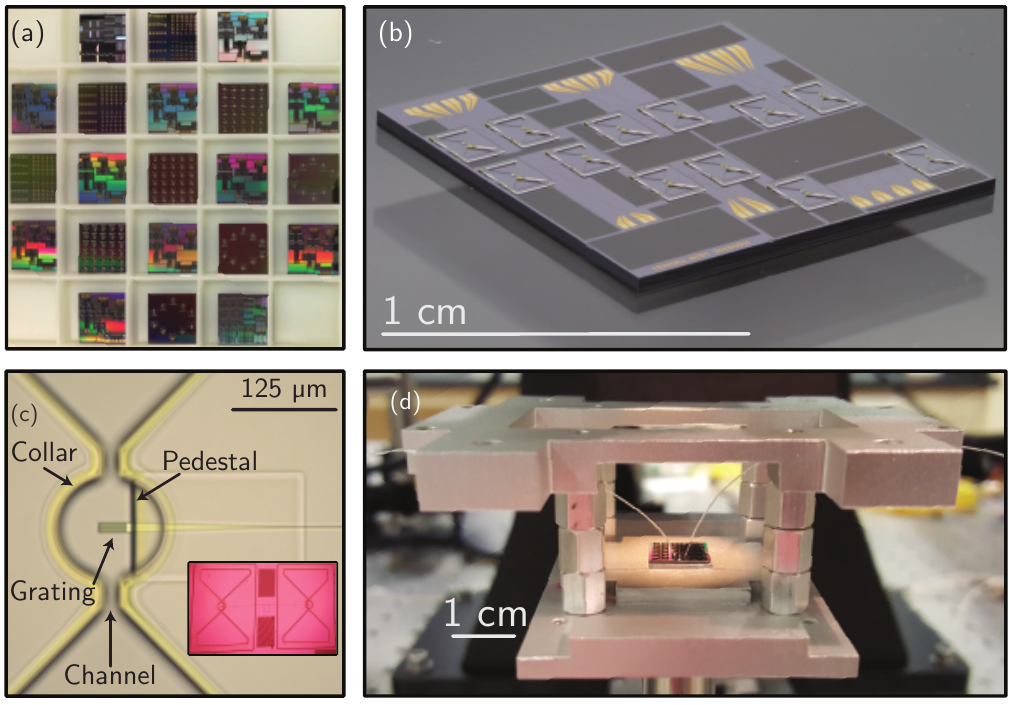}
	\caption{\label{fig:chipsAndPackaging} (a) A fabricated and diced wafer. (b) A focus-stacked image of a single die (1 cm$^2$). (c) Optical microscope image of the fiber collar and pedestal above a SiN grating. (d) The fully packaged die prepared for mounting in a cryostat.}
\end{figure}

After patterning the SU-8 collar layer, the wafer is diced, and the chips are ready for fiber packaging. A fully processed and diced wafer is shown in Fig. \ref{fig:chipsAndPackaging}(a). It can be seen that multiple die types have been fabricated. A close up of one particular die is shown in Fig. \ref{fig:chipsAndPackaging}(b). The SU-8 pedestal and collar are shown in Fig. \ref{fig:chipsAndPackaging}(c). A grating coupled to a tapered waveguide is also visible, and the inset shows two such fiber collars with a ring coupled to a bus between. In Fig. \ref{fig:chipsAndPackaging}(c) one can see that the fiber collar has two channels on the sides of the fiber. The low viscosity of the epoxy enables it to run through these channels into the ellipse housing the fiber. Upon filling the entire volume, the epoxy is UV-cured. After placing all fibers, an additional amount of epoxy can be delivered to further strengthen the assembly. A fully packaged chip suitable for cryogenic applications is shown in Fig. \ref{fig:chipsAndPackaging}(d).

This packaging approach is suitable for a wide range of operating temperatures. With Dymax OP-4-20632 cryogenic epoxy, we demonstrated cooling of a packaged sample down to $<$ 1 K, incurring an additional 1.5 dB insertion loss per coupler. This sample has been thermally cycled twice with the same performance. Additional thermal cycles have not been attempted. With Norland Optical Adhesive, we demonstrated heating up to 160 \degree C. Upon heating to 120 \degree C\ an additional 1 dB insertion loss per coupler was incurred. The sample was held at 120 \degree C\ for 16 hours with no additional insertion loss. It was then raised to 160 \degree C\, incurring an additional 1 dB insertion loss per coupler. Such packaged chips have been observed to maintain stable coupling for weeks, and the fully encased fibers are protected from the environment to enable exceptional longevity. 

We wish to use our waveguide devices to perform spectroscopy in outdoor environments, and also for quantum optical experiments with superconductors in cryogenic environments. Toward these ends, the packaging approach presented here has several strengths. The simplicity renders it useful in many contexts, as it does not depend on KOH wet etch steps to form v-grooves \cite{Cohen2013,Beyer2015}, which can be difficult from a process-integration standpoint, and it is not dependent on a two-chip assembly \cite{Barwicz2014}. Alignment of the collar relative to the grating is achieved lithographically, so little optimization is required. Fiber couplers can be placed anywhere on the die for higher density, not just the periphery, as in many v-groove \cite{Cohen2013,Beyer2015} and two-chip approaches \cite{Barwicz2014}. Packaging can be highly automated as the placement of a fiber in a collar and the epoxy delivery is well-suited to robotics with machine vision. 

\section{\label{sec:nonlinear}Applications in nonlinear optics}
SiN waveguides provide an excellent platform for nonlinear optics due to their high nonlinear index (10x larger than silica) \cite{makl2015} and low absorption from visible to infrared wavelengths. Recent progress \cite{chbo2014,oksa2011} has revealed opportunities to use group-velocity-dispersion (GVD) engineered devices in conjunction with high-power ultrafast lasers to generate spectra suitable for applications such as spectroscopy \cite{cone2016}, time and frequency metrology \cite{dijo2000,ha2006,bahe2009,di2010}, and novel optical sources.

\begin{figure} %[t] %[htb]
	\includegraphics{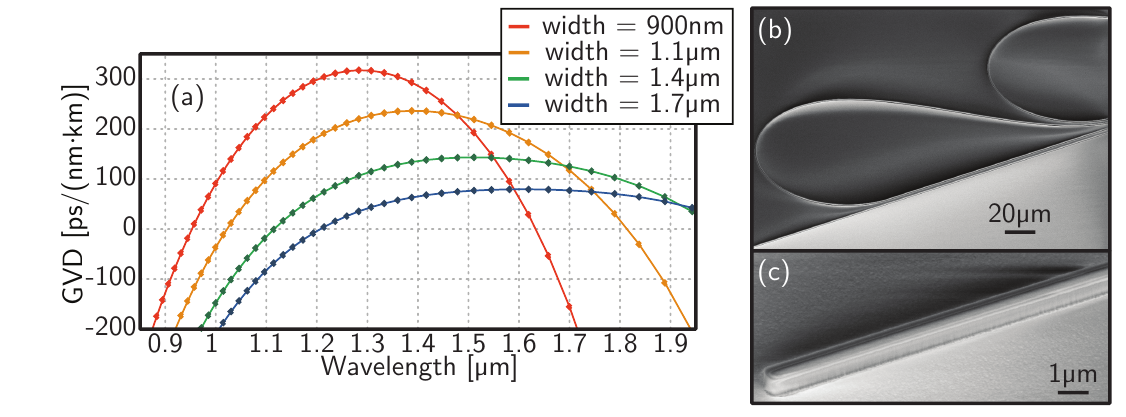}
	\caption{\label{fig:nonlinear1} (a) GVD curves for 700 nm thick SiN waveguides with air over-cladding for different waveguide widths. (b) Scanning electron micrograph of paperclip waveguide structures for measuring propagation losses. (c) Tapered waveguide for edge-coupling.}
\end{figure}
Finite-difference frequency domain calculations show near-zero GVD is achieved for wavelengths in the near-IR by engineering SiN waveguides with thicknesses above 600 nm. For the present study, we investigated waveguides with 700 nm thickness. The GVD curves for these waveguides are shown in Fig. \ref{fig:nonlinear1}(a) for various waveguide widths. These calculations assume thermal SiO$_2$ as an undercladding, air as an overcladding, and employ the SiN index data shown in Fig. \ref{fig:layers}(b).  Waveguides with widths $>$ 1.4 $\mu$m have very flat, near-zero anomalous GVD with one zero crossing at short wavelengths.  These waveguides can be used for octave-spanning supercontinuum generation which is useful for self-referenced frequency combs \cite{dijo2000,ha2006,di2010}. Slightly narrower waveguides have anomalous GVD with two zero crossings. As the waveguide width is reduced below 1 $\mu$m, the anomalous GVD region becomes narrower and drops more steeply to negative values. These waveguides generate narrow, flat-top spectra. Such spectra improve the signal-to-noise ratio for spectroscopic applications targeting a specific narrow frequency band.

Figure \ref{fig:nonlinear1}(b) and (c) show scanning electron micrographs of fabricated waveguides. In addition to the processing described in Sec. \ref{sec:basicPassives}, these waveguides incurred a resist reflow at 150 \degree C\ for five minutes before etching. The WSi, SU-8, and overcladding layers were omitted for these devices. Figure \ref{fig:nonlinear1}(b) shows a waveguide in a paperclip configuration \cite{orta2012} to measure propagation losses.  These measurements give 1.8 dB/cm propagation loss, an improvement over the loss-ring measurement of the thin waveguides presented in Sec. \ref{sec:basicPassives} (2.5 dB/cm). The loss has not been observed to depend on intensity, indicating that multiphoton absorption is a smaller contribution than sidewall scattering even for femtosecond pulses with 6 kW peak power. Figure \ref{fig:nonlinear1}(c) shows a waveguide with a tapered tip for edge coupling, which was used for these nonlinear optics experiments. Insertion loss of 5.2 dB/facet was measured. For spectroscopy applications in the field or metrology applications in a system, the fiber packaging approach presented in Sec. \ref{sec:packaging} can be utilized.

\begin{figure} %[t] %[htb]
	\includegraphics{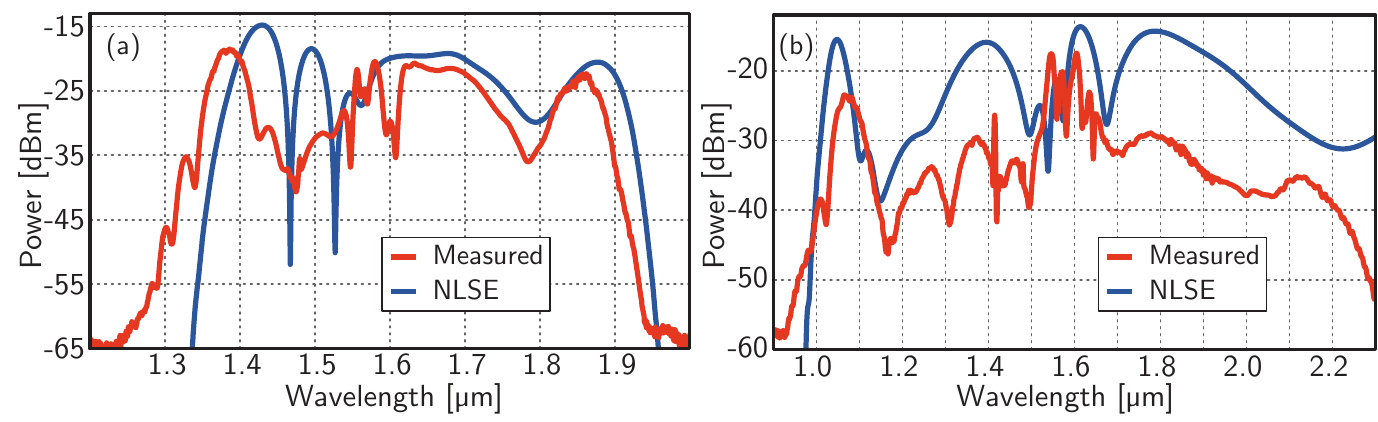}
	\caption{\label{fig:nonlinear3} (a) Supercontinuum generation from a 960 nm wide waveguide. (b) Supercontinuum from a 2.1 $\mu$m wide waveguide. Light is coupled out of the waveguide into a single-mode fiber and into an optical spectrum analyzer where the power is measured. For both of these measurements, the peak laser power was 5.25 kW, the repetition rate was 200 MHz, and the average power was 63 mW.}
\end{figure}

For investigation of nonlinear spectral broadening, a high-power, femtosecond 1550 nm fiber frequency comb with up to 17 kW peak power and 60 fs pulse duration was used as the laser source \cite{side2015}. The source was TE-polarized and coupled into the waveguide tapers with a lensed fiber. Output spectra were recorded as a function of incident power with an optical spectrum analyzer. The measured outputs of 900 nm- and 1900 nm-wide waveguides are presented in Figs. \ref{fig:nonlinear3}(a) and (b), respectively. Also shown on these plots are the theoretical predictions using the GVD curves of Fig. \ref{fig:nonlinear1}(a) in the nonlinear Schr\"odinger equation (NLSE). The 900 nm device generates a supercontinuum that stays within 20 dB over the wavelength range from 1350 nm to 1850 nm and sharply drops beyond this spectral range. This relatively flat spectrum is suitable for spectroscopy wherein a flat spectrum provides a correspondingly flat response function with a uniform signal-to-noise ratio. 

In contrast to the spectrum of Fig. \ref{fig:nonlinear3}(a) the spectrum of Fig. \ref{fig:nonlinear3}(b) shows broad supercontinuum generation from a 2100 nm-wide waveguide. This supercontinuum covers the spectrum from 1050 to 2200 nm when pumped with comb peak power of 5.25 kW. Such a coherent comb spectrum is suitable for self-referenced frequency combs. Additionally, due to the design of the GVD zero crossings at 1.1 $\mu$m and 2.2 $\mu$m, the optical power is concentrated at these wavelengths which will be utilized in the difference-frequency generation process.

These examples of spectral tailoring for spectroscopic applications as well as frequency metrology applications show the utility of this deposited-waveguide fabrication process for nonlinear optics. 

\section{\label{sec:quantum}Integration with single-photon detectors}

For quantum optical applications including metrology, linear-optical quantum computing, and quantum networks, single-photon detection with near-unity efficiency is paramount. For systems utilizing many components, an integrated photonic environment with fiber optics and on-chip devices operating at telecom wavelengths is the most promising route to scaling. Within this context, SNSPDs are a promising candidate for single-photon detection with high efficiency and low timing jitter. The fabrication of SNSPDs on waveguide devices \cite{Cavalier2011,Sprengers2011,Tanner2012,Pernice2012b,Atikian2014a,Akhlaghi2015,Kahl2015,Beyer2015,Sahin2015,Najafi2015,Schuck2016} has been described in Sec. \ref{sec:basicPassives}. 

The detection mechanism of an SNSPD depends on the absorption of a photon locally breaking Cooper pairs in the superconductor, leading to the production of a region of normal metal, referred to as a ``hot spot''. With sufficient bias current, the absorption of a photon by the SNSPD will result in a voltage pulse \cite{gook2001,nata2012,liyo2013,keya2009,jafi2012,cahe2015,cahe2013,mabi2009,dima2008,yake2007,fija2003,Marsili2013}. The required wire dimensions for this process depend on the wavelength of the light and the material properties. Typical SNSPD materials are NbN \cite{Goltsman2001} or WSi \cite{Baek2011}. WSi devices typically have a lower critical temperature and higher timing jitter than NbN, but at present have higher yield and higher detection efficiency. Here we explore WSi SNSPDs.

For detection of 1550 nm photons, typical WSi wires are 130 nm wide and 4 nm thick \cite{Marsili2013}. In wires much wider than this, the hot spot created via photon absorption is not large enough to span the width of the wire and result in a voltage pulse, even when the wire is biased very close to its critical current. However, using the lithographic system available to us, we cannot achieve such narrow features. To circumvent this issue, we experimented with varying the film thickness with the intention of achieving similar photon detection efficiency with wider wires. 

\begin{figure} %[t] %[htb]
	\includegraphics{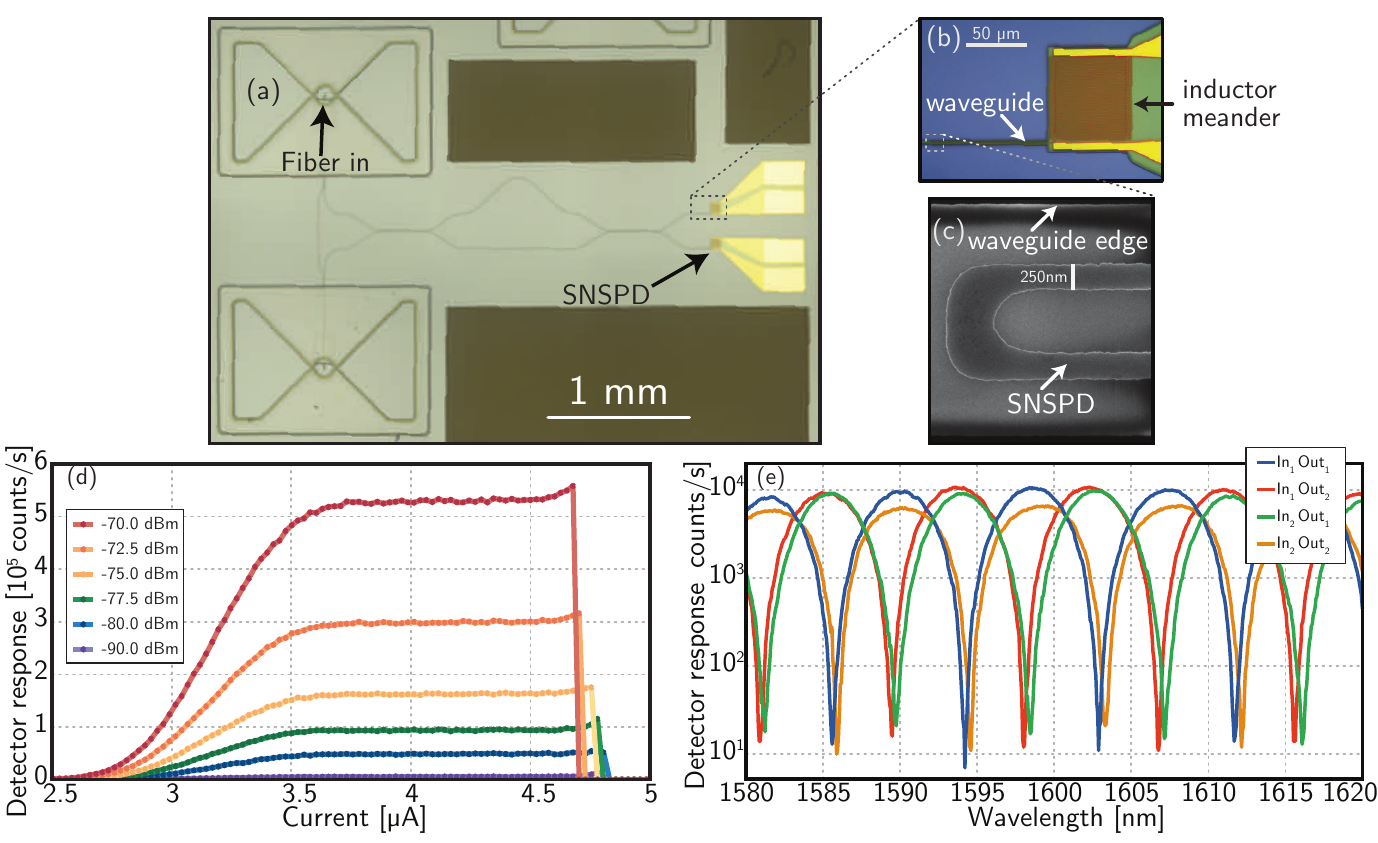}
	\caption{\label{fig:mzi} (a) Optical microscope image of an MZI. Between the two input ports, a waveguide tapping 10\% of the light before entering the MZI is used to optimize fiber alignment at room temperature. (b) Optical microscope image of a waveguide-integrated detector. (c) Scanning electron micrograph of the tip of a waveguide-integrated SNSPD. (d) Response of a single detector versus bias current for various optical powers. (e) Measurements of waveguide-integrated SNSPDs at the output ports of an MZI.}
\end{figure}

We incorporated these SNSPDs into several photonic devices. Figure \ref{fig:mzi}(a) shows an optical micrograph of an MZI with two fiber input ports and WSi SNSPDs at the output ports. Figure \ref{fig:mzi}(b) shows an optical micrograph of the waveguide-integrated SNSPD, and Fig. \ref{fig:mzi}(c) shows a scanning electron micrograph of the photolithographically patterned nanowire with 250 nm width. 
 
The response of a waveguide-integrated WSi SNSPD with 250 nm width and 3.5 nm (85\% the typical thickness \cite{Marsili2013}) to light at 1570 nm is shown in Fig. \ref{fig:mzi}(d) for different input powers at the grating. For these measurements, the device was operated at 740 mK. Not only do we find that these wider, thinner wires are capable of detecting photons of this energy, we observe a broad plateau as a function of bias current, indicating saturated internal quantum detection efficiency. We note that thinning of our SNSPD material suppresses the critical temperature and current, leading to reduced performance, limiting the extent to which thinning the material can relax the lithographic requirements. A thorough study of SNSPD performance across the parameter space of width and thickness will be the subject of future work. 

The optical transmission spectra from each of the two MZI input ports to each of the two detectors is shown in Fig. \ref{fig:mzi}(e), and nearly 30 dB extinction is observed. The data in Fig. \ref{fig:mzi}(d) was acquired from one of these detectors, and their performance was nearly identical. The minor offset of the spectra along the wavelength axis is due to imprecise timing alignment between the sweep of the laser and the SNSPD data acquisition, resulting in a slight inaccuracy in the recorded wavelength.

\begin{figure} %[t] %[htb]
	\includegraphics{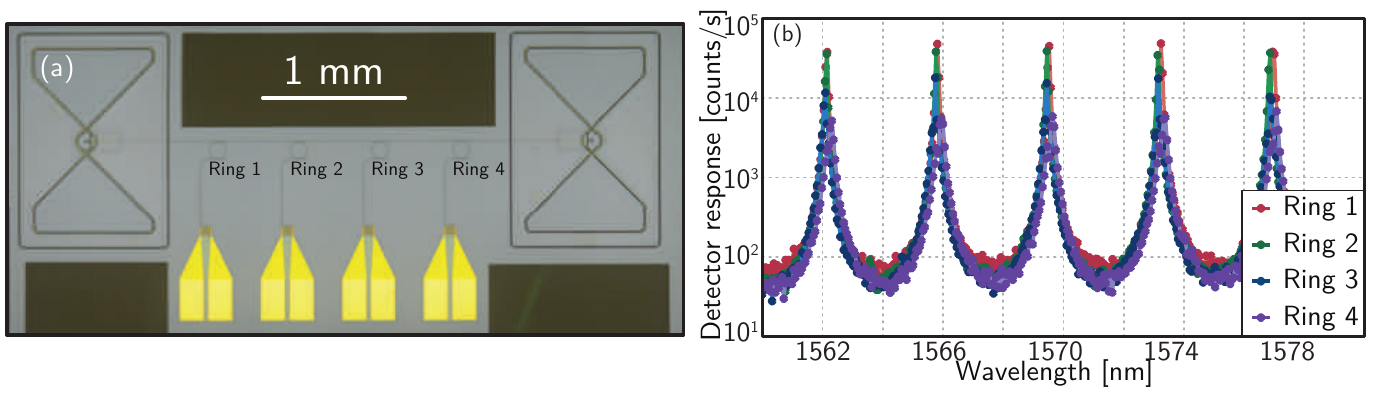}
	\caption{\label{fig:fOF} (a) Optical microscope image of four ring filters dropping to detectors. (b) Spectrum of the device.}
\end{figure}
Finally, we integrated the SNSPDs at the drop ports of four first-order filters. An optical micrograph of the optoelectronic devices is shown in Fig. \ref{fig:fOF}(a). The ring design used for this device is the same as that shown in Fig. \ref{fig:passives}(c), except with a drop port as well as a through port and a 650 nm ring-bus gap on both ports. The detector response as a function of wavelength is shown in Fig. \ref{fig:fOF}(b) for the detectors at all four drop ports. Extinction of $>$ 20 dB is observed for all four detectors. We note that the yield of these detectors fabricated from WSi using i-line lithography was quite high, and it was not necessary to test multiple MZI devices or filter banks to find one with all detectors working. Such a fabrication process is scalable to large systems without the need for pick-and-place device assembly methods \cite{Najafi2015}. 
	
\section{\label{sec:summary}Summary and Outlook}
We have presented an approach to fabrication and packaging of integrated photonic devices that utilizes waveguide and detector layers deposited at near-ambient temperature. All lithography was performed with a 365 nm i-line stepper. We have shown low-loss waveguides, high-$Q$ ring resonators, critically coupled ring resonators, 50/50 beam splitters, MZIs, and a robust fiber-packaging system. We have further explored the utility of this process by demonstrating spectral tailoring and supercontinuum generation of a fiber-generated frequency comb. Finally, we demonstrated the applicability of this fabrication process to quantum-optical systems by integrating SNSPDs with MZIs and channel-dropping filters.

Because all layers in this process are deposited at near-ambient temperature, and because the fiber packaging approach is compatible with a wide variety of substrates, devices, and processes, this fabrication process is compatible with BEOL incorporation on fully processed CMOS wafers or wafers of integrated photonic devices fabricated by other foundries. Additionally, because we have only made use of deposited films for the waveguide and detector layers, it is possible to extend this platform by vertically stacking many layers of waveguides and detectors using interlayer waveguide couplers to construct complex, 3D photonic systems.

Regarding the utility of this scalable and nimble fabrication infrastructure, we note that the cost of all equipment used to perform the fabrication of these devices is quite low. The PECVD tool costs \$400K, and can also serve as the RIE tool. The sputtering tool also costs \$400K. The i-line stepper is \$3.2M. Thus, the entire capital investment for this equipment could be as low as \$4M, making this approach accessible for start-ups, small businesses, academics, and innovators in developing countries. If added to an existing foundry, the additional capital investment would be a small fraction of the total cost of the foundry. Based on these results, we are of the opinion that as integrated photonics moves to a foundry process model, innovation per dollar can be maximized through investment in nimble and diverse processes serving a wider range of application spaces. 
\newline
\\[2pt]
\noindent This is a contribution of NIST, an agency of the US government, not subject to copyright. Product disclaimer: Any mention of commercial products is for information only; it does not imply recommendation or endorsement by NIST.

\bibliography{SiN,sin_new}

\end{document}